\documentclass[12pt]{iopart}

\usepackage{epsfig}

\newcommand{\be}{\begin{equation}}
\newcommand{\ee}{\end{equation}}
\newcommand{\bea}{\begin{eqnarray}}
\newcommand{\eea}{\end{eqnarray}}

\begin{document}

\title{Charmed hadron signals of partonic medium}

\author{O.~Linnyk$^1$, E.~L.~Bratkovskaya$^1$, W.~Cassing$^2$}

\address{$^1$
Frankfurt Institute for Advanced Studies, %
 Ruth-Moufang-Str. 1, %
 60438 Frankfurt am Main, %
 Germany}%
\address{$^2$
Institut f\"ur Theoretische Physik, %
  Universit\"at Giessen, %
  Heinrich--Buff--Ring 16, %
  35392 Giessen, %
  Germany}%

\ead{linnyk@fias.uni-frankfurt.de}


\begin{abstract}
We present a short review of our results on the collectivity and the
suppression pattern of charmed mesons in heavy-ion collisions
based on the microscopic  Hadron-String Dynamics (HSD) transport approach
for different scenarios of charm interactions with the surrounding matter --
the `comover' dissociation by mesons with further recreation by
$D-\bar D$ channels and `pre-hadronic' interaction scenarios.
While at SPS energies the  hadronic `comover' absorption
scenario is found to be compatible with the experimental data, the
dynamics of $c, \bar{c}$ quarks at RHIC are dominated by partonic or
`pre-hadronic' interactions in the strongly coupled quark-gluon plasma
stage and cannot be modeled by pure `hadronic' interactions.
We find that the collective flow of charm in the purely hadronic
scenario appears compatible with the data at SPS
energies but underestimates the data at top RHIC energies. Thus, the large
elliptic flow $v_2$ of $D$-mesons and the low $R_{AA}(p_T)$ of $J/\Psi$
seen experimentally at RHIC have to be attributed to early interactions of
non-hadronic degrees of freedom. Simultaneously, we observe that
non-hadronic interactions are mandatory in order to describe the
narrowing of the $J/\Psi$ rapidity distribution from $pp$ to central
$Au+Au$ collisions at the top RHIC energy.  We
demonstrate additionally that the strong quenching of high-$p_T$
$J/\Psi$'s in central $Au+Au$ collisions indicates that a
fraction of final $J/\Psi$ mesons is created by a coalescence mechanism
close to the phase boundary.

\end{abstract}



\section{Introduction}
\vspace{-2mm}

The phase transition from partonic degrees of freedom to interacting
hadrons, as occurring in relativistic nucleus-nucleus collisions, is a
central topic of modern high-energy physics. The main difficulty in the
interpretation of the data is that the information about the initial
strongly interacting quark-gluon-plasma (sQGP) stage of matter can be
obtained only indirectly from the measurement of hadronic observables;
it might be strongly distorted by the hadronization process and final
state interactions of the hadrons. In order to reliably subtract the
hadronic contribution from the sQGP signal carried by charmed mesons
(as well as hadronic final-state interactions)
we apply a microscopic transport approach.


Our study is based on the Hadron-String-Dynamics~\cite{Cass99} transport
approach, which simulates the time-dependent `hadronic environment' for
open charm mesons and charmonia in heavy-ion collisions rather
well~\cite{brat03}.  Upon being produced in initial $N+N$ collisions,
charmonia in HSD are absorbed on baryons (normal nuclear absorption) as
well as on `comoving' mesons by the process $c\bar{c}+meson\to
D+\bar{D}, D^*+\bar{D}, D^*+\bar{D}^* $ etc. Note that the latter
interactions lead also to a recreation
of charmonia via the inverse recombination process. The $J/\Psi,
\chi_c, \Psi^\prime$ formation cross sections by open charm mesons or
the `comover' dissociation cross sections are not well known. In HSD a
simple 2-body transition model is employed that allows to implement the
backward reactions uniquely by exploiting detailed balance for each
individual channel. The details of the model are described in the review
~\cite{charm.review}. In the current proceedings we concentrate on
the results and the comparison to data.

In the hadronic comover dissociation and recombination scenario only
formed comoving mesons participate in the dissociation of charm or $D\bar D$
recombination reactions. Indeed, in the default HSD all newly produced
hadrons (by string fragmentation)  have a formation time of $\tau_F
\approx$0.8 fm/c $\approx 1/\Lambda_{QCD}$ in their rest frame and do
not interact hadronically during the `partonic' propagation. Furthermore,
hadronization is inhibited, if the energy density -- in the local rest
frame -- is above 1 GeV/fm$^3$, which roughly corresponds to the energy
density for QGP formation in equilibrium. Being hard probes, $c\bar c$
pairs are created in the early stage of the collision, while the
comoving mesons are formed at a later stage.

In order to simulate partonic interaction effects we have included
explicit interactions of pre-hadrons with the charmed
mesons~\cite{Olena.RHIC.2}. In HSD a {\it pre-hadron} is defined as a
state with the quantum numbers of a hadron being at time $t$ under
`formation' ($t<\tau_F \gamma$) or in a cell with  local energy
density above $\approx 1$~GeV/fm$^3$ (cf. Ref.~\cite{charm.review}).
Since the cross sections of elastic $c\bar c$  scattering on `pre-mesons'
and `pre-baryons' are unknown, we have adjusted them to RHIC data
on $J/\Psi$ suppression (cf. Ref.~\cite{charm.review}).
It has to be stressed that further explicit partonic degrees of
freedom, i.e. gluons and their mutual interactions and gluon
interactions with quarks/antiquarks, have not been taken into account
explicitly so far.
\begin{figure}
\centerline{\psfig{file=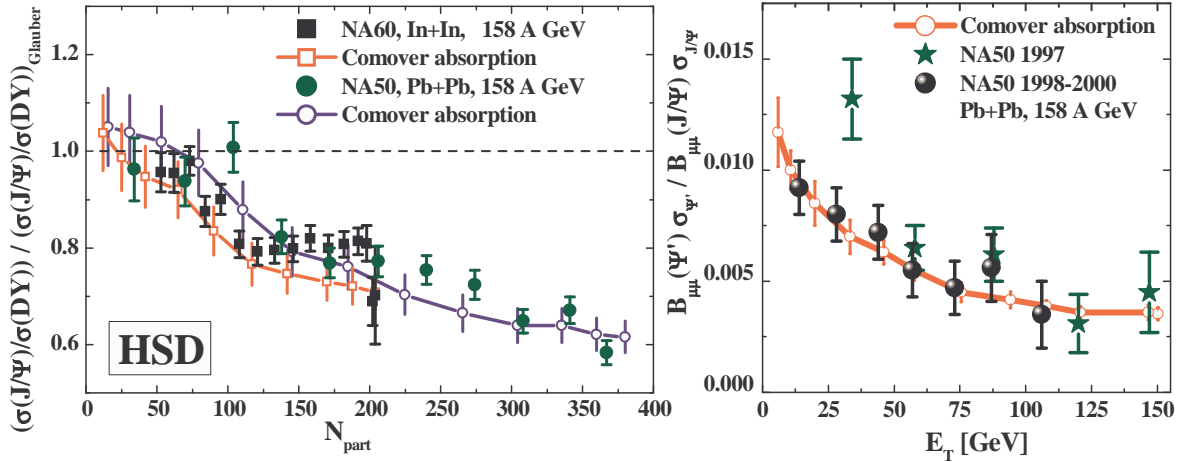,width=\textwidth}} \caption{{\bf Left panel: } The ratio
$B_{\mu\mu}\sigma(J/\Psi)/\sigma(DY)$ as a function of the number of participants $N_{part}$ in
In+In (red line with open squares) and Pb+Pb reactions (blue line with open circles) at 158
A$\cdot$GeV relative to the normal nuclear absorption given by the straight black line. The full
dots and squares denote the respective data from the NA50 and NA60 Collaborations. The model
calculations reflect the hadronic comover absorption scenario. {\bf Right panel: }
The  $\Psi^\prime$ to
$J/\Psi$ ratio as a function of the transverse energy $E_T$ for Pb+Pb at 160 A$\cdot$GeV. The full dots
and stars denote the respective data from the NA50 Collaboration \protect\cite{NA50PsiPrime}. The
HSD result~\protect\cite{Olena.SPS} for the comover absorption scenario is shown by the red line.
The figures are taken from Ref.~\protect\cite{Olena.SPS}.}
\label{figure1}
\end{figure}

\section{Comparison to data}
\vspace{-2mm}

We start our comparison to data with the charmonium production and
suppression at SPS energies within the default `hadronic comover' scenario.
As found in Ref.~\cite{Olena.SPS}, the comover absorption
model performs well with respect to all data sets at SPS.
Indeed, the extra suppression of charmonia by comovers, seen in
Fig.~\ref{figure1},  matches the $J/\Psi$ suppression in In+In and Pb+Pb
as well as the $\Psi^\prime$ to $J/\Psi$ ratio rather well. One may
conclude that the comover absorption model so far cannot be ruled out
on the basis of the available data sets from the SPS within error bars.

\begin{figure}
  \hspace{0.05\textwidth}
  \begin{minipage}[b]{0.5\textwidth}
\centerline{    \psfig{file=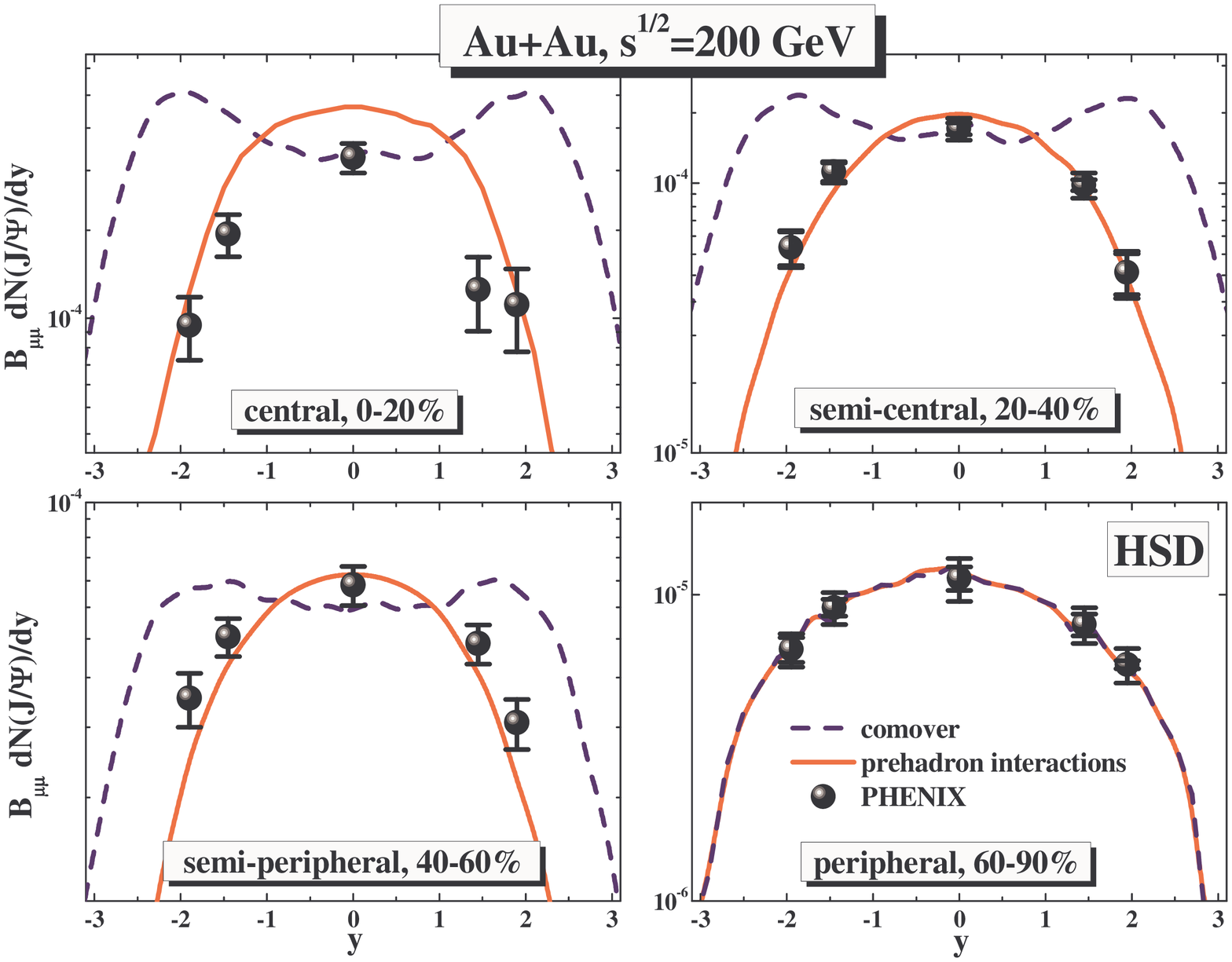,width=1.1\textwidth}}
  \end{minipage}
  \hspace{-0.05\textwidth}
  \begin{minipage}[b]{0.5\textwidth}
    \caption{The rapidity distribution of
$J/\Psi$'s for different centralities in the `comover' approach (dashed
blue lines) and the `comover' approach with additional pre-hadronic
interactions of charm (solid red lines). The full dots show the
respective data from the PHENIX
Collaboration~\protect\cite{PHENIXNov06}. The calculated lines have
been smoothed by a spline algorithm.  The reactions are Au+Au at
$\sqrt{s}=200$~GeV. The figure is taken from Ref.~\protect\cite{Olena.RHIC.2}. 
}
        \label{y1}
  \end{minipage}
\end{figure}

Further information may be gained from the $J/\Psi$ rapidity
distributions in  Au+Au collisions at RHIC. The latter distribution is
shown in Fig.~\ref{y1} in comparison to the PHENIX data in the standard
`comover' scenario (dashed blue lines) and the `comover' model
including pre-hadronic interactions of charm (solid red lines).
Whereas for peripheral reactions the additional early interactions
practically play no role, the pre-hadron elastic scatterings lead to a
dynamical narrowing of the $J/\Psi$ rapidity distribution with the
centrality of the collision (roughly in line with the data). In the
standard `comover' model an opposite trend is seen: here the
interactions of charmonia with formed hadrons produce a dip in the
rapidity distribution at $y\approx0$ which increases with centrality
since the density of formed hadrons increases accordingly around
mid-rapidity. Since the total number of produced $c{\bar c}$ pairs is
the same (for the respective centrality class) and detailed balance is
incorporated in the reaction rates, we find a surplus of $J/\Psi$ at
forward rapidities. The net result is a broadening of the $J/\Psi$
rapidity distribution with centrality in the purely hadronic scenario
opposite to the trend observed in experiment.  Consequently, the PHENIX
data on $J/\Psi$ suppression indicate the presence and important impact
of pre-hadronic or partonic interactions in the early charm dynamics.

\begin{figure}
\centerline{\psfig{file=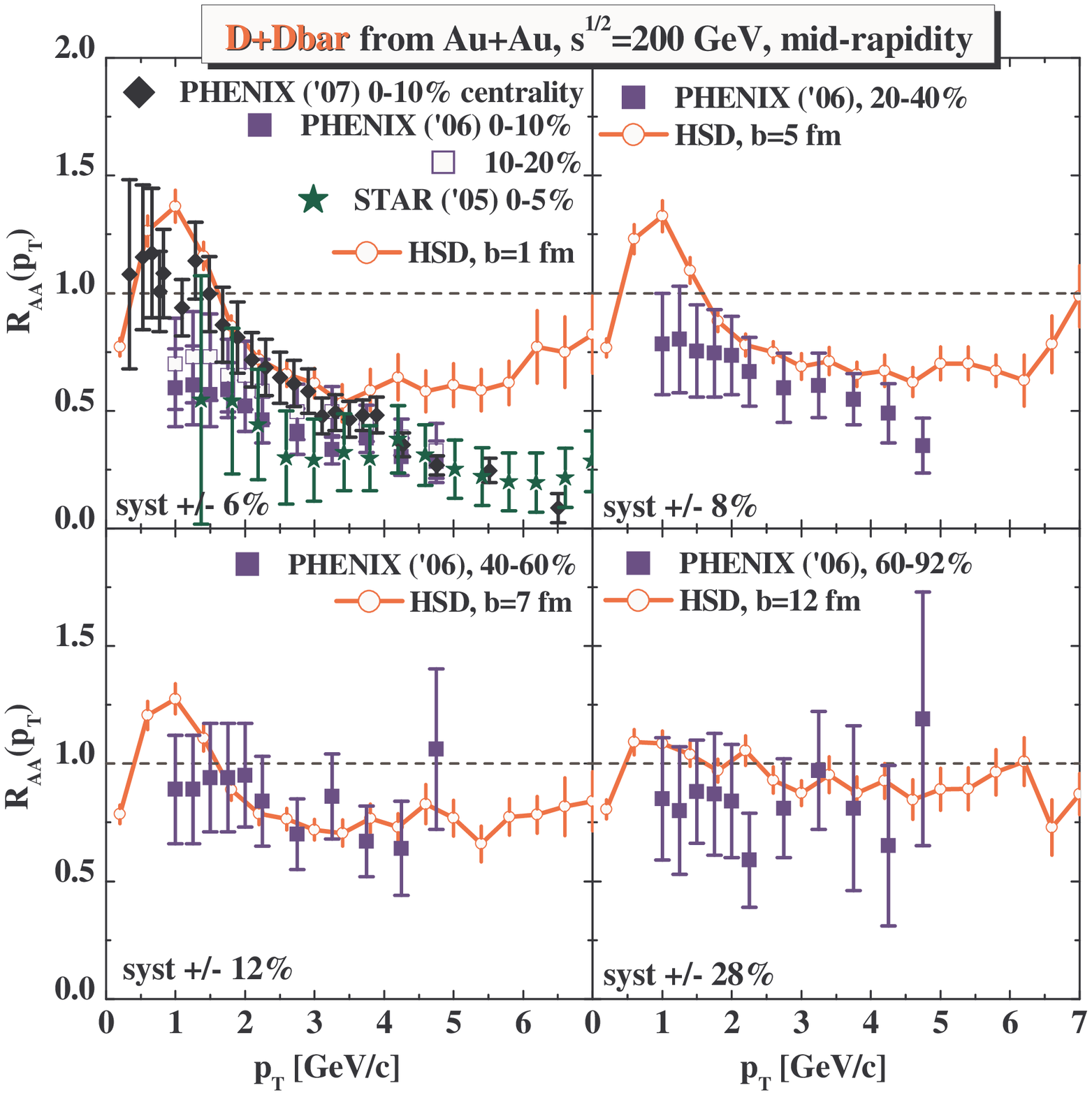,width=0.495\textwidth}\psfig{file=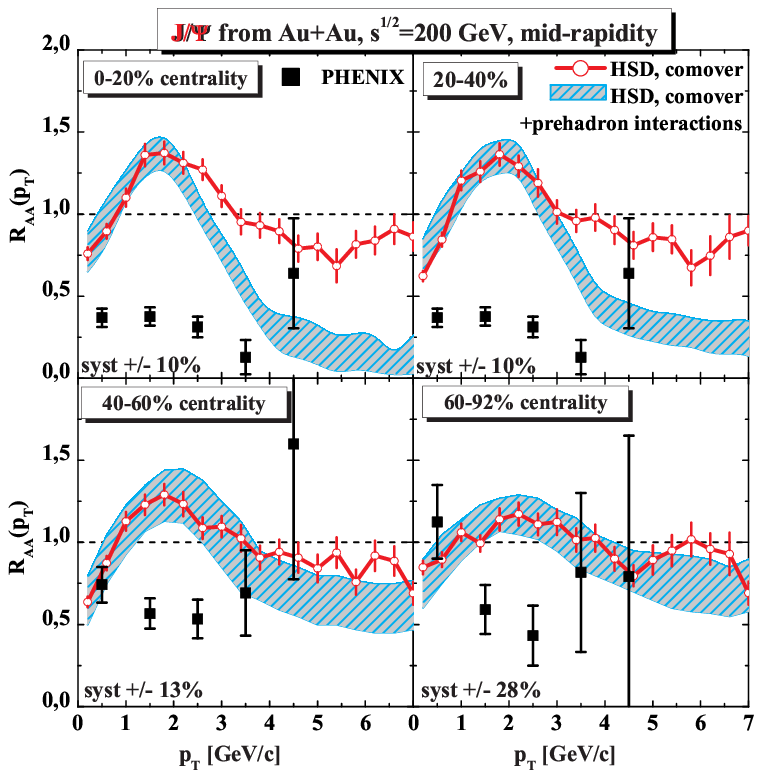,width=0.495\textwidth}}
\caption{{\bf Left panel: } HSD predictions for the ratio of the final to the initial
(i.e. at the production point) transverse momentum spectra of $D+\bar
D$-mesons (solid lines with open dots, color: blue) from Au + Au
collisions at $\sqrt{s}=200$ GeV for $b=1, 5, 7$ and 12 fm at
mid-rapidity as predicted in Ref.~\protect\cite{brat05}. The
PHENIX data from Ref.~\protect\cite{LuizdaSilva:2006vw} (denoted
'06) and from Ref.~\protect\cite{PHENIXv2D} (denoted '07) on
$R_{AA}$ of non-photonic electrons as well as the STAR
data from Ref.~\protect\cite{Bielcik:2005wu} have been added later.
{\bf Right panel: } The same observable for $J/\Psi$. The solid
lines show the calculations~\protect\cite{brat05} in the default HSD, while the grey
dashed bands represent the results from the extended version of the HSD
comover approach, in which pre-hadron interactions are taken into
account. The preliminary PHENIX data for $R_{AA}(J/\Psi)$ from
Ref.~\protect\cite{PHENIXNov06} have been added later.
The figures are taken from Ref.~\protect\cite{charm.review}.} \label{D-Raa-PT}
\end{figure}

A significant suppression of high transverse momentum hadrons in Au+Au
collisions compared to $pp$ is observed at RHIC energies of
$\sqrt{s} =$ 200 GeV
and is attributed to the energy loss of highly energetic particles in a
hot colored medium (QGP). In order to quantify {\em the effect of
hadronic final state interactions}, we show in the left panel of
Fig.~\ref{D-Raa-PT} the HSD predictions from Ref.       ~\cite{brat05} for the
ratio of the final to the initial transverse $p_T$ spectra of $D+\bar
D$-mesons from Au + Au collisions at $\sqrt{s}=200$ GeV. HSD predicts
an enhancement of $D, \bar{D}$ mesons at low momenta with a maximum at
$p_T \approx $ 1 GeV/c and a relative suppression for $p_T > $ 2 GeV/c.
These effects increase with the centrality of the Au+Au collision. We
note that the maxima in the ratios disappear when switching off the
rescattering with mesons in the transport approach. Thus a collective
acceleration of the $D+\bar{D}$ mesons occurs also via elastic
scattering with mesons. As shown in Fig.~\ref{D-Raa-PT} the suppression
seen by PHENIX may well be explained by hadronic comover interactions
up to transverse momenta about 4 GeV/c. Only for higher $p_T$ a clear
signal for parton energy loss -- either gluon bremsstrahlung or parton
elastic scattering -- may be extracted from comparison to the data!

The suppression pattern $R_{AA}(J/\Psi)$ from  HSD (in the comover
scenario) is quite analogous to that of $D$-mesons showing a slight
maximum for transverse momenta of $\sim$ 2 GeV/c and a steady decrease
for higher $p_T$. In the right panel of Fig.~\ref{D-Raa-PT} the
predictions (from Ref.~\cite{brat05}) for the ratio of the final to the
initial transverse $p_T$ spectra of $J/\Psi$-mesons as well as new
calculations in the extended version of the comover approach (grey
dashed bands) from Au + Au collisions at $\sqrt{s}=200$~GeV  are
displayed. The preliminary PHENIX data for $R_{AA}(J/\Psi)$
from Ref.~\cite{PHENIXNov06} -- added later -- show a substantially
different pattern, especially for non-peripheral interactions. The
strong suppression for low $p_T$ $J/\Psi$ mesons seen experimentally
suggests that not primordial $J/\Psi$'s are accelerated during the
dynamical evolution but that at least a part of initially formed
$J/\Psi$'s are dissolved and recreated later, e.g. by $c\bar{c}$
coalescence. We stress that the reformation of charmonia in the
hadronic phase (by $D + \bar{D}$ etc.) carries the flow from the
$D$-mesons and thus does not lead to suppression at small $p_T$ as seen
experimentally. This observation supports the idea that part of the
charmonia are produced in the hadronization process!

A further possible way to disentangle hadronic from partonic dynamics
is the elliptic flow $v_2(y,p_T)$. In Ref.~\cite{Olena.RHIC.2} we
compared the HSD result for $v_2 (J/\Psi)$ at SPS in the purely
hadronic `comover' scenario to the data for $v_2 (J/\Psi)$ of the NA60
collaboration for In+In collisions~\cite{NA602007}. The agreement found
between the theory and the data indicates that -- in line with the
reproduction of the $J/\Psi$ suppression data~\cite{Olena.SPS} (cf.
Fig.~\ref{figure1}) -- the (low) $v_2(J/\Psi)$ does not point towards
additional strong partonic interactions at SPS energies.

\begin{figure}
  \begin{minipage}[b]{0.5\textwidth}
    \psfig{file=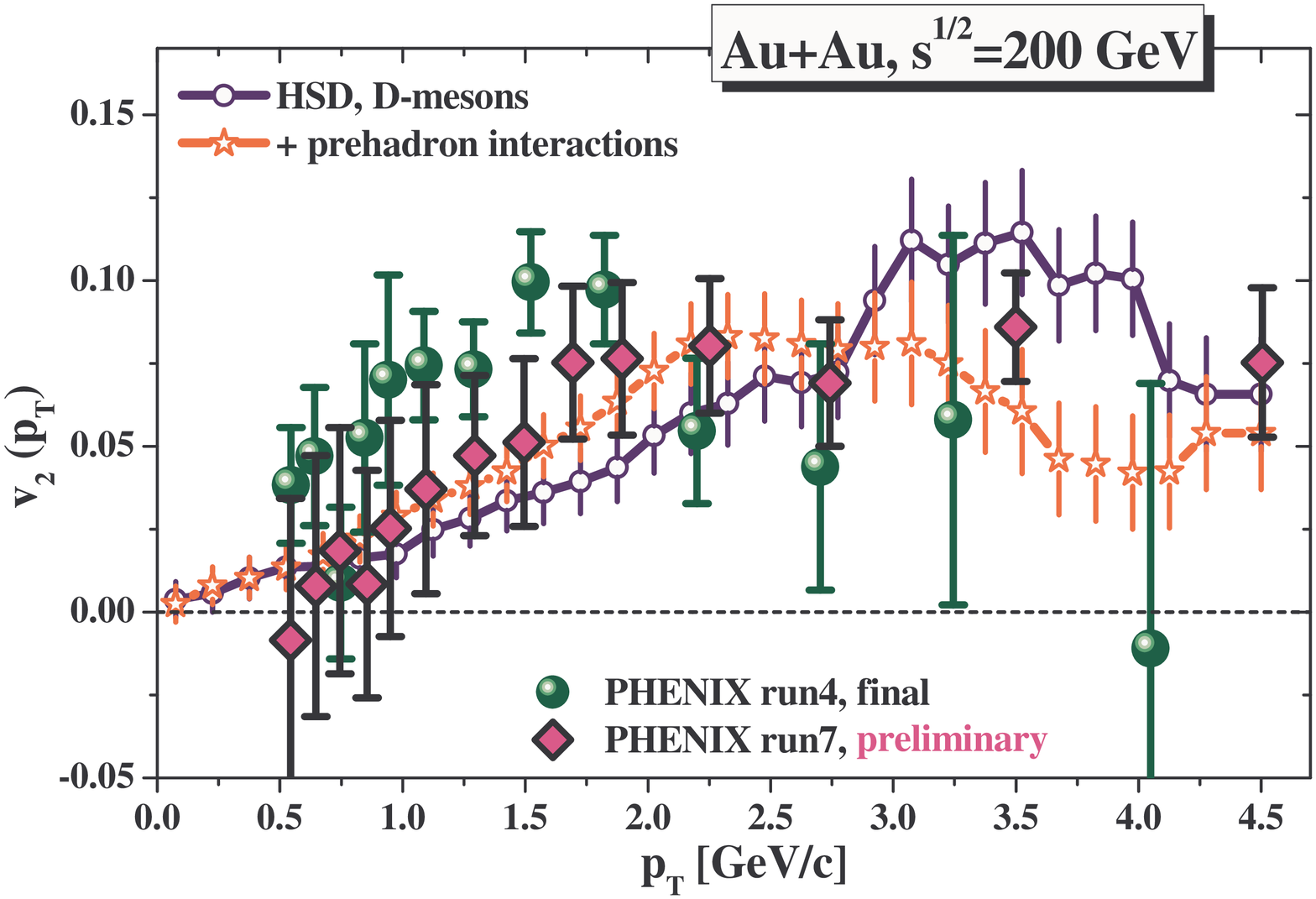,width=1.3\textwidth}
  \end{minipage}
  \begin{minipage}[b]{0.5\textwidth}
    \caption{Elliptic flow of $D$-mesons
produced in $Au+Au$ collisions at $\sqrt{s}=200$~GeV as a function of $p_T$ from HSD (solid blue
line with open circles) in comparison to the PHENIX data~\protect\cite{PHENIXv2D} on $v_2$ of
non-photonic electrons. The red line with open stars shows the HSD result for the $v_2$ of
$D$-mesons when including additionally pre-hadronic charm interactions.
The figure is taken from Ref.~\protect\cite{Olena.RHIC.2}. \vspace{12pt}}
    \label{Dv2}
  \end{minipage}
\end{figure}

The situation is, however, different for the collective flow of
$D$-mesons at top RHIC energies. In Fig.~\ref{Dv2} one sees that the
elliptic flow of $D$-mesons is clearly underestimated in the default
(purely hadronic) HSD model (cf. Ref.~\cite{brat05}). Only when including
pre-hadronic charm interactions, the elliptic flow moderately
increases, but still stays below the PHENIX data. We thus conclude that
the modeling of charm interactions by pre-hadronic interactions
accounts for part of the non-hadronic generation of the $v_2$, but does
not provide enough interaction strength in the early phase of the
collision.

Since a large fraction of $J/\Psi$'s in central Au+Au collisions at
RHIC are created by $D\!+\!{\bar D}$ recombination, the elliptic flow
of $J/\Psi$'s  obtained from HSD in the comover (purely hadronic) case
is comparatively small, too. The accuracy of the preliminary PHENIX
data so far does not allow for a differentiation between the different
model predictions (cf. Ref.~\cite{charm.review}).

\section{Conclusions}
\vspace{-2mm}

The above findings suggest that the charmonium dynamics in heavy-ion
reactions is dominantly driven by hadronic interactions in the SPS
energy regime. Since energy densities above 1 GeV/fm$^3$ are reached in
central nucleus-nucleus collisions at 158 A$\cdot$GeV, our observation
indicates that hadronic correlators (with quantum numbers of the
familiar hadrons) still persist above the critical energy density for
the formation of a QGP.

In contrast, the study of the formation and suppression dynamics of
charmonia within the HSD transport approach for Au+Au reactions at the
top RHIC energy has demonstrated that the hadronic `comover absorption
and recreation model' fails severely at $\sqrt{s}$=200 GeV. This is
found in the $J/\Psi$ rapidity distribution, in the differential
elliptic flow of $J/\Psi$ and the charmonium nuclear modification
factor $R_{AA}$ as a function of transverse momentum $p_T$. Only when
including pre-hadronic degrees in the early charm reaction dynamics,
the general suppression pattern of charmonia may be reasonably
described; though, the elliptic flow $v_2$ is still (slightly)
underestimated. On the other hand, $R_{AA}(p_T)$ for $J/\Psi$ mesons
cannot be described appropriately in the comover approach even when incorporating
the early pre-hadron interactions. The latter observable  indicates that
at least part of the final $J/\Psi$'s are created by coalescence of
$c\bar{c}$ pairs in the hadronization phase. Our analysis demonstrates
that the dynamics of $c, \bar{c}$ quarks in heavy-ion reactions at RHIC
energies are dominated by partonic interactions in the sQGP.

\section*{References}
\vspace{-2mm}



\end{document}